# Cross-sectional helium irradiation reveals interface-controlled bubble evolution in Cr/CrAlSiN multilayer coatings on zirconium alloys

Renda Wang [a,h], Xue Bai [a,c], Xueliang Pei [a], Sijie Liu [a,h], Chunfan Liu [a,h], Ping Yu [b], Bingsheng Li [d,1*], Nabil Daghbouj [e,2*], Tomas Polcar [e,i], Fanping Meng [a,f], Fangfang Ge [a,f,g,3*], Qing Huang [a,f,g]

[a] Zhejiang Key Laboratory of Data-Driven High-Safety Energy Materials and Applications, Ningbo Key Laboratory of Special Energy Materials and Chemistry, Ningbo Institute of Materials Technology and Engineering, Chinese Academy of Sciences, Ningbo 315201, China.

[b] School of Electronics and Information Engineering, Ningbo University of Technology, Ningbo 315211, China

[c] College of Materials Science and Chemical Engineering, Harbin Engineering University, Harbin 150001, China

[d] State Key Laboratory for Environment-friendly Energy Materials, Southwest University of Science and Technology, Mianyang 621010, China.

[e] Department of Control Engineering, Faculty of Electrical Engineering, Czech Technical University in Prague, Technická 2, 160 00, Prague 6, Czech Republic.

[f] Qianwan Institute of CNITECH, Ningbo 315336, China.

[g] Advanced Energy Science and Technology Guangdong Laboratory, Huizhou 516000, China.

[h] University of Chinese Academy of Sciences, Beijing 100049, China.

[i] School of Engineering, University of Southampton, Southampton SO17 1BJ, United Kingdom

---

[1]* Corresponding author (gefangfang@nimte.ac.cn). Tel: 86-574-86685035; Fax: 86-574-86685159

[2]* Corresponding author (libingshengmvp@163.com).

[3]* Corresponding author (daghbnab@fel.cvut.cz).

**Abstract**

The irradiation stability of Cr-based protective coatings on zirconium alloys is critical for the development of accident-tolerant fuel claddings. However, conventional surface irradiation often produces shallow, nonuniform damage, obscuring interfacial behavior. In this study, we perform cross-sectional $He^{2+}$ irradiation to directly examine the interfacial response and He bubble evolution across Cr monolayer and Cr/CrAlSiN multilayer coatings on Zr substrates. Irradiation was carried out at 500 °C and 750 °C to doses of 2–3 dpa, enabling a direct comparison of temperature-dependent microstructural evolution. In the Cr monolayer, He implantation produced a homogeneous distribution of nanoscale bubbles throughout the damaged region and large cavities at the Cr/Zr interface, indicating severe Kirkendall-type voiding and interfacial decohesion at elevated temperature. In contrast, the Cr/CrAlSiN multilayer exhibited a periodically modulated bubble distribution, with bubble fragmentation and transformation into nanoscale platelets at CrAlSiN interfaces. A N-enriched Zr(N) interlayer formed spontaneously at the CrAlSiN/Zr interface, effectively suppressing bubble accumulation and interdiffusion. The nanochannel interfaces acted as He sinks and diffusion barriers, enhancing interfacial bonding and mitigating swelling. This work demonstrates that cross-sectional ion irradiation is a powerful approach for probing interfacial stability in multilayer systems, offering new insights into He-defect interactions and radiation tolerance engineering at buried interfaces. The findings highlight the potential of Cr/CrAlSiN multilayers as advanced coating architectures for high-temperature nuclear environments.

# 1. Introduction

Nuclear energy — clean, sustainable, and cost-effective — offers one of the surest routes to easing environmental degradation and the global energy crunch. Yet since the Fukushima Daiichi accident, the imperative of nuclear safety has moved to the forefront of public concern [1]. Advanced reactor structural materials must endure a perfect storm of extreme temperatures, high stress, and intense neutron flux [2]. To mitigate both the rate and total generation of heat during severe accident scenarios, such as reactivity insertion accidents (RIA) and loss-of-coolant accidents (LOCA), various strategies involving accident-tolerant fuel (ATF) cladding materials have been proposed and are currently being advanced worldwide [3]. The development of ATF claddings generally follows two primary approaches: (i) the introduction of entirely new material classes, including $SiC_f/SiC$ composites [4], FeCrAl alloys [5], and high-entropy alloys [6]; and (ii) the enhancement of zirconium (Zr) alloy claddings through surface coatings, which retain the inherent benefits of zirconium alloys, such as low thermal neutron absorption, resistance to irradiation-induced swelling and creep, and favorable mechanical properties [7], while adding coatings that significantly improve high-temperature steam oxidation resistance [8]. Clearly, Zr-based alloys with protective coatings present a practical ATF cladding solution in the near term, pending the resolution of limitations associated with entirely new cladding materials. Currently, numerous surface coatings with thicknesses ranging from a few to several tens of micrometers are under development. These coatings include: (i) metallic coatings such as Cr [9, 10] and FeCrAl [11]; and (ii) ceramic coatings such as $Al_2O_3$ [12], CrAlSi [13], SiC [14], ZrN [15], and CrAlSiN [16]. Despite advances in oxidation-resistant coatings, their response to helium (He) irradiation, rather than only neutron damage, has not been sufficiently optimized. This is crucial because helium-induced degradation can dominate mechanical failure modes in reactor environments.

When exposed to energetic neutrons, countless atoms in the material are displaced from their lattice sites, giving rise to a complex population of radiation-induced defects. These include simple point defects, such as vacancies and interstitials, as well as more extended defect structures, including dislocation loops, stacking-fault tetrahedra, and ultimately voids [17-19]. Simultaneously, neutron-induced transmutation reactions generate insoluble gases, primarily He atoms. Due to its extremely low solubility in materials, He readily associates with vacancies, forming stable clusters that grow into pressurized bubbles [20-22]. These bubbles tend to accumulate preferentially along microstructural features, such as dislocations, precipitates, and grain boundaries, where they can serve as additional nucleation sites for void formation [23, 24]. The combined evolution of these defects and gas-filled bubbles progressively alters the material's microstructure, ultimately impairing its mechanical integrity. This manifests as radiation-induced hardening, reduced ductility, and embrittlement, which are among the most critical challenges in ensuring the long-term performance and safety of structural materials in nuclear environments [25, 26]. Developing materials capable of withstanding the damaging influence of He remains one of the central challenges in nuclear engineering.

Conventional coatings can delay oxidation and swelling, but they typically lack mechanisms for dynamically managing helium generation, transport, and release during service. Studies have shown that under He-ion irradiation, nanochannel CrN films exhibit superior structural stability compared to compact CrN films, owing to the efficient release of helium through nanochannels formed during physical vapor deposition (PVD) [27]. Building on the same concept, nanochannel W [28] and CrMoTaWV high entropy alloy (HEA) [29] PVD films have also been developed. Because these nanochannels promote He out-diffusion, the growth rate of irradiation-induced fuzz in nanochannel W and HEA films is reduced by factors of 1.9 and 2.3, respectively, relative to their bulk counterparts [28, 29]. However, a key limitation is that the nanochannels, typically aligned with the columnar grain boundaries in PVD coatings, extend straight through the entire coating thickness. These long-range diffusion pathways may increase susceptibility to mechanical failure and corrosion, or oxidation [30, 31].

Therefore, it is necessary to engineer architectures that retain the He-mitigating benefits of nanochannels while avoiding continuous pathways that compromise coating integrity. To address this issue, an alternative irradiation-tolerant strategy was previously proposed: the incorporation of amorphous-like textures as uniformly distributed, low-density channels in CrAlSiN coatings. In this architecture, approximately 40% of implanted He was released after annealing at 800 °C for 30 minutes [16]. This approach aligns with a broader strategy that relies on microstructures engineered with efficient defect sinks, particularly interfaces, which can trap He atoms and enhance defect recombination before extensive damage develops. Interfaces such as multilayer interfaces, and heterointerfaces provide excess free volume, altered bonding environments, and chemical heterogeneity that lower the formation and migration energies of radiation-induced point defects [32-34]. As a result, He can migrate preferentially toward these regions instead of aggregating in the matrix, while vacancies and interstitials recombine more easily, delaying bubble nucleation, void growth, swelling, hardening, and embrittlement [35, 36]. Building on this principle, it was further demonstrated that inserting periodic nanochannel CrAlSiN layers into a Cr coating significantly reduces swelling and enhances oxidation resistance after Fe-ion irradiation at 400 °C, with the nanochannel layers acting as elastic, unsaturated "sponges" [37]. Moreover, the Cr/CrAlSiN multilayer architecture enhances both mechanical performance and oxidation resistance by inhibiting dislocation motion, columnar sliding, and the rapid diffusion of oxidative/corrosive species [37]. However, the thickness of the amorphous layer is a critical parameter, as excessive thickness may lead to coating brittleness.

Moreover, the Cr/Zr interface exhibits limited thermal stability at elevated temperatures, where the formation of a $ZrCr_2$ Laves phase has been widely reported. During high-temperature exposure, Cr atoms diffuse into the Zr substrate and precipitate as $ZrCr_2$, particularly during the β to α phase transformation of Zr, as demonstrated in previous studies [38-41]. This interfacial reaction not only alters the local microstructure but also induces stresses and degrades the coating adhesion, ultimately compromising the performance of Cr-coated Zr claddings under reactor conditions. To mitigate this interdiffusion issue, the introduction of an effective diffusion barrier between the Cr coating and Zr substrate has been proposed [42]. Several refractory metals, such as Mo [43, 44] and Nb [45], have been investigated as potential barrier layers due to their high melting points and limited mutual solubility with both Cr and Zr. In addition, other studies have suggested using CrN interlayers [46, 47] or medium-entropy alloy (MEA) coatings [48] to suppress interdiffusion and improve interfacial stability.

However, despite these efforts, the combined effects of irradiation and high temperature on interfacial evolution remain insufficiently understood. To gain deeper insight into the mechanisms governing irradiation resistance, this work investigates how periodically inserted nanochannel CrAlSiN layers within a Cr matrix regulate helium bubble nucleation, growth, and transport under high-temperature $He^{2+}$ irradiation. By comparing the microstructural evolution at 500 °C and 750 °C, we elucidate the role of multilayer interfaces in dynamic helium accommodation, defect management, and interfacial stability under reactor-relevant conditions. In ATF concepts, the protective coating is directly exposed to neutron irradiation, primary knock-on atoms, and irradiation-induced defect production, whereas the Zr-based cladding primarily serves as a structural substrate. Consequently, irradiation effects in the coating materials themselves, rather than in bulk Zr, are of primary relevance. Although neutron-induced helium production in Zr alloys under conventional light-water reactor conditions is negligible, the response of advanced coating materials to gas implantation remains an important subject, particularly for multi-component systems with distinct defect energetics and gas-trapping behavior. The He implantation conditions employed in this study were intentionally selected to enable direct observation of bubble nucleation and evolution within the temporal constraints of in situ transmission electron microscopy. The applied fluence, therefore, represents an accelerated testing condition designed to probe gas–defect interactions and comparative irradiation responses of different coating chemistries, rather than to reproduce the exact helium inventory expected in current fission reactors. Recent studies have demonstrated that local strain fields, arising from interfaces, grain boundaries, phase mismatch, or irradiation-induced swelling, play a decisive role in defect energetics and transport. Both experimental and simulation-based investigations have shown that strain can modify defect formation energies, migration barriers, and clustering behavior, thereby strongly influencing He bubble evolution and radiation tolerance. In particular, combined nanobeam electron diffraction measurements and atomistic simulations by Daghbouj et al. [32, 35, 49] have established a direct correlation between local strain gradients and defect diffusion pathways in irradiated materials. These findings motivate the present focus on multilayer architectures, where interfaces and strain heterogeneities may synergistically enhance helium and radiation-induced damage management.

## 2. Experimental details

Multilayer Cr/CrAlSiN and monolayer Cr coatings, with the thickness of ~10 μm were deposited on Zr-alloy (composition: Nb 1.10 wt.%, Sn 1.10 wt.%, Fe 0.11 wt.%, O 0.13 wt.%, Zr bal.) coupons with dimensions of 15 mm × 15 mm × 1.5 mm. The coatings were fabricated using magnetron sputtering in a chamber equipped with two cathodes. A high-purity Cr target (99.9 % in purity) and a CrAlSi target (Cr 61.2 at.%, Al 25.4 at.%, Si 13.4 at.%), both 10 cm in diameter, were mounted on the respective cathodes. The CrAlSi target was powered by a mid-frequency (MF) pulsed supply (Advanced Energy Pinnacle® Plus +5/5) in the direct-current (DC) mode, while the Cr target was driven by a radio-frequency (RF) power supply (Comdel CV-1000, 81 MHz) in parallel to another DC power supply. The base pressure before deposition was ~2.0 × $10^{-5}$ Pa, and the working pressure was ~0.7 Pa. A detailed description of the deposition process of the coatings can be found in the reference [50]. The multilayer Cr/CrAlSiN coating consisted of alternating layers of Cr (~900 nm) and amorphous CrAlSiN (~150 nm) for nine cycles, with the outermost layer being CrAlSiN.

The coatings were irradiated from cross-section in their as-deposited state, and cross-sectional TEM lamellae were subsequently prepared after irradiation using focused ion beam (FIB) milling to avoid any modification of irradiation-induced defects prior to analysis. The He ion-irradiation experiments were performed at the 320 kV Multidisciplinary Research Platform for Highly Charged Ions of the Institute of Modern Physics, Chinese Academy of Sciences (CAS). The details of the irradiation condition are provided in **Table 1**. The ion energy of 300 keV was selected to ensure that the projected range of $He^{2+}$ ions overlapped with the full thickness of the multilayer structure when irradiated from the cross-section. This allowed comparable damage levels to develop across both the Cr and CrAlSiN layers, which is essential for evaluating interface-governed bubble evolution. The fluence of $1 \times 10^{17}$ ions/cm² was chosen to generate He concentrations of several atomic percent, comparable to levels expected in structural components subjected to long-term neutron irradiation in advanced reactors, and sufficient to promote observable bubble nucleation, growth, and coalescence under TEM. The irradiation temperatures of 500 °C and 750 °C were selected to simulate normal and off-normal service conditions. These temperatures enable the assessment of thermally activated helium migration and trapping, and provide a basis for comparing the evolution of bubbles under different defect mobility regimes. Surface irradiation typically produces only a shallow damage layer with pronounced depth-dependent gradients, making it difficult to achieve uniform irradiation across all sublayers of a multilayer coating. To circumvent this limitation, the samples were irradiated from the cross-section. Additionally, while heavy and self-ion irradiation can generate high displacement damage (tens to hundreds of dpa) comparable to neutron exposure, the penetration depth of MeV heavy ions is generally less than 1 μm, insufficient to reach the Cr/Zr interface in through-thickness irradiation. Therefore, cross-sectional irradiation was the most effective strategy for verifying interface stability.

**Fig. 1(a)** shows the SEM image with the incident direction of the He ions and the FIB lift-out position of the irradiated samples. The EBSD image in **Fig.1(b)** shows the various grain orientations of Cr layers and the tiny CrAlSiN layers that are represented with black color. The corresponding displacements per atom (dpa) and ion concentration as a function of depth were simulated using the SRIM-2013 program in Kinchin-Pease mode [51], and the results are presented in **Fig. 1(c)**. The peak He concentration and damage levels were calculated to be ~6.0 at.% and ~2.3-2.8 dpa, respectively. The input atomic densities of Cr and CrAlSiN layers were $8.338 \times 10^{22}$ atoms/cm$^3$ and $7.285 \times 10^{22}$ atoms/cm$^3$, respectively. For the displacement threshold energy ($E_d$) that depends on crystallographic orientation and bonding environment, after comprehensively referring to relevant references and standard manuals, it was set to 40 eV for Cr, 40 eV for Al, 13 eV for Si, and 25 eV for N [52-54]. The He peak concentration in CrAlSi is deeper than in Cr, as shown in **Fig.1c**. All irradiation conditions were kept identical for the different coatings to ensure that the observed differences in microstructural evolution arise from intrinsic material responses rather than variations in irradiation parameters. When analyzing, He bubbles in irradiated samples; at least 100 bubbles were counted in each region to ensure statistical reliability. Swelling was calculated from the measured bubble size and density within each region to provide a quantitative metric of volumetric expansion.

**Table 1** Irradiation conditions of the Cr/CrAlSiN coatings.

| Sample | Irradiated face | Ion | Energy (keV) | Temperature (°C) | Fluence (ions/cm$^2$) |
|---|---|---|---|---|---|
| S1 | Cross-section | He$^{2+}$ | 300 | 750 | $1 \times 10^{17}$ |
| S2 | | | | 500 | |

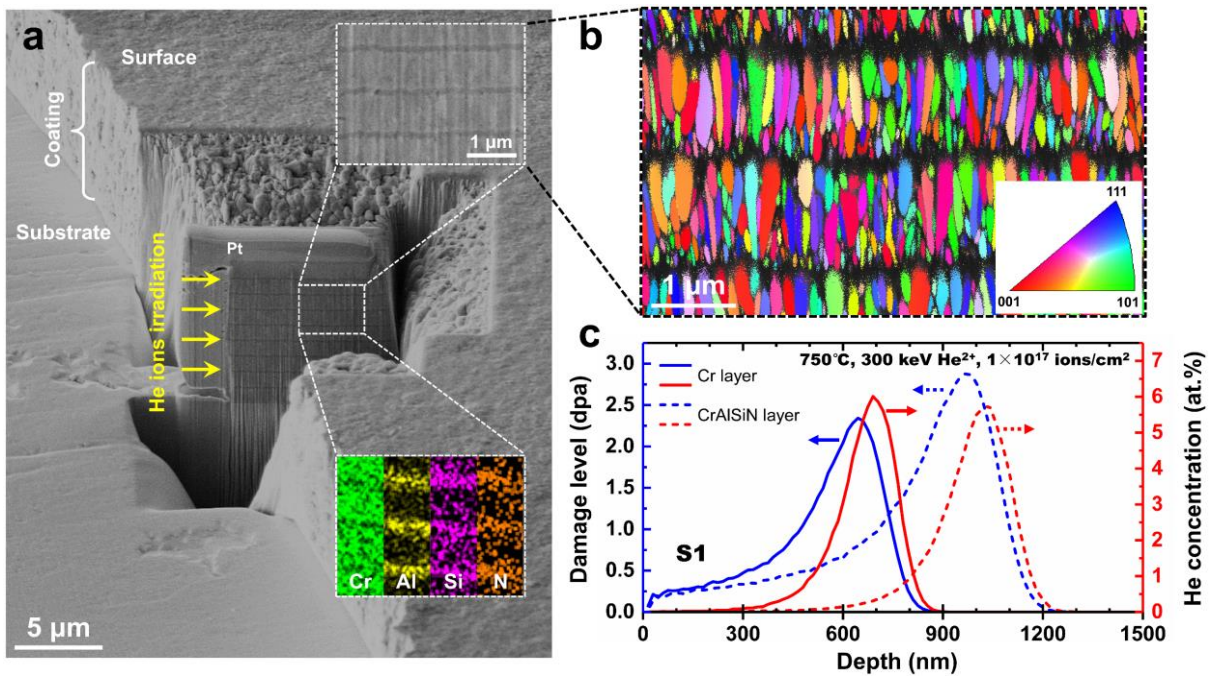

**Fig. 1** (a) SEM micrograph showing the cross-sectional He$^{2+}$ irradiation direction and FIB milling position of the Cr/CrAlSiN-coated Zr alloy. (b) An EBSD orientation map highlighting the interrupted columnar growth structure of the coating. (c) Depth profiles of displacement per atom (dpa) and He ion concentration calculated using the SRIM code for the Cr/CrAlSiN coatings irradiated with 300 keV He$^{2+}$ ions at a fluence of $1 \times 10^{17}$ ions/cm$^2$.

The coating morphologies were examined by a Dual Beam Focused Ion Beam (DB-FIB) workstation (FEI, Helios-G4-CX) equipped with an electron backscatter diffraction (EBSD) detector (Bruker, CrystAlign). Transmission electron microscopy (TEM) was performed using an FEI Talos F200X operated at 200 kV. Cross-sectional TEM specimens were prepared via the FIB lift-out. As shown in **Fig. 2**, the S2 sample was additionally shaped into a TEM specimen, welded onto a heating holder, and subjected to *in situ* annealing up to 750 °C, under TEM observation. The lamella thickness, estimated to be approximately 80–100 nm, was measured from the electron images acquired by the built-in FIB system. This thickness range ensures sufficient electron transparency while allowing partial stress relaxation through free surfaces, thereby minimizing artefacts associated with constrained deformation or artificial defect accumulation. *In situ* TEM observations were performed using a defocus value of approximately 1.4 μm.

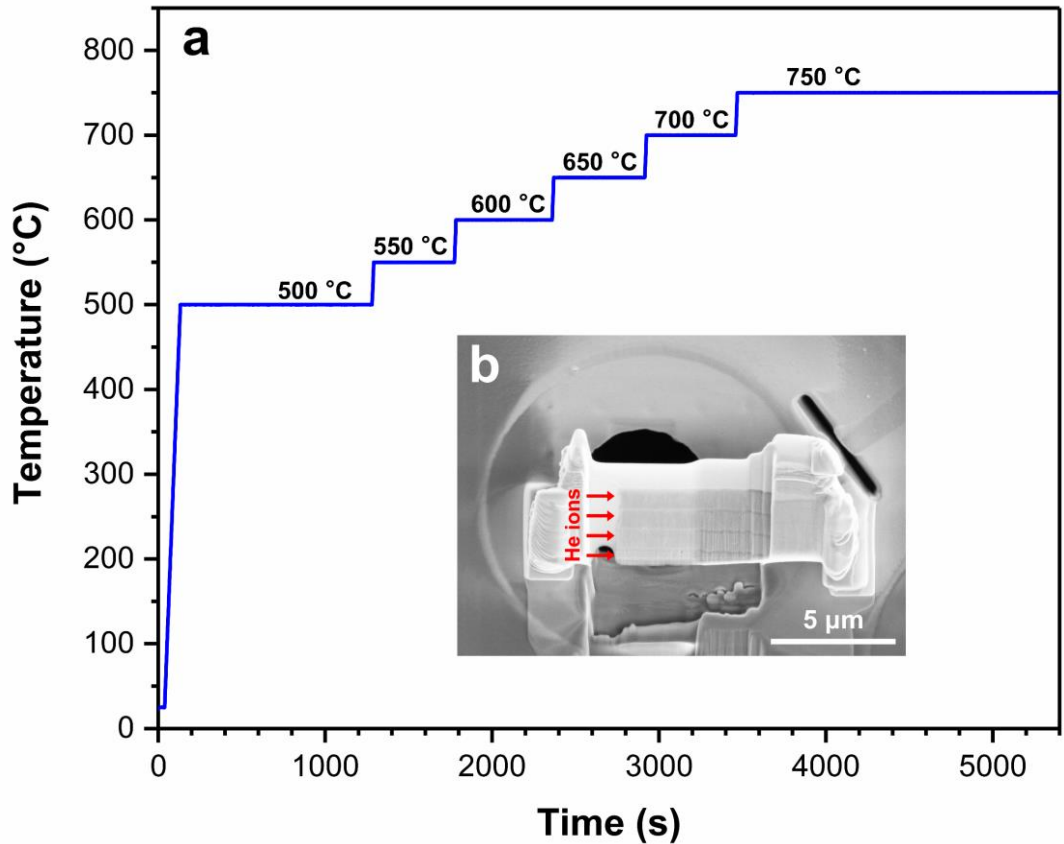

**Fig. 2** (a) TEM in situ heating profile (temperature vs. time) and (b) SEM image of the in-situ heating sample mounted on the holder.

During in situ heating experiments, thermal expansion mismatch between the specimen, weld material, and TEM holder may introduce additional stresses. However, given the nanoscale thickness of the lamellae and the gradual heating protocol with sufficient dwell times, such stresses are expected to be largely relaxed through free surfaces. No evidence of bending, cracking, or mechanically induced contrast was observed, suggesting that the reported microstructural evolution is primarily governed by thermally activated defect processes rather than extrinsic mechanical effects.

## 3. Results

### 3.1 Microstructure of the as-deposited coatings

The growth structure of the Cr/CrAlSiN coating, as shown in **Fig. 1(a)**, was studied by SEM and EBSD examinations. The multilayer Cr/CrAlSiN coating consisted of alternating layers of Cr (~900 nm) and amorphous CrAlSiN (~150 nm) for nine cycles, with the outermost layer being CrAlSiN. The bilayers combined tightly without any obvious delamination, although the introduction of CrAlSiN layers interrupted the growth of the Cr grains. As the EBSD map shows, the columnar grains in each Cr layer will re-nucleate on the previous CrAlSiN layer, achieving a fine columnar structure with the columnar width of less than ~70 nm as well as an increase of GB density, being different from the single-layer structure Cr coatings, which are generally composed of coarse columns (hundreds of nanometers in width). Moreover, the individual CrAlSiN layer is featured with a special amorphous-like structure, containing uniformly distributed nanochannels, and more details can be found elsewhere [16].

*3.2    Microstructure of the coatings after the He ions irradiation at 750 °C*

After the irradiation at 750 °C, **Fig. 3(a)** shows a zone of uniformly sized He bubbles (~$1.76 \times 10^{23}$ m$^{-3}$, ~4.0 nm in diameter) that are homogeneously distributed throughout the Cr coating, as depicted in **Fig. 3(b)-(e)**, which exhibit nearly identical characteristics of the He bubbles. The square-like or faceted morphology of some He bubbles is attributed to elastic and crystallographic anisotropy in the bcc Cr lattice. Under high-temperature irradiation, bubble shapes may become faceted along low-index crystallographic planes to minimize interfacial and elastic energy, consistent with previous observations in cubic metals [55, 56]. Such faceting occurs preferentially when bubble size exceeds a critical radius and when local elastic constraints and crystallographic orientation favor anisotropic interfacial energy minimization, explaining why only a subset of bubbles exhibits square-like morphologies. In contrast to the Cr monolayer, the Cr/CrAlSiN multilayer coating exhibits pronounced layer-dependent variations in He bubble size and number density, indicating a dynamic redistribution process governed by interfaces and local chemical and structural heterogeneities (**Fig. 4** and **Fig. S1**). The Cr layers (**Fig. 4(b)-(d)**) contain He bubbles of varying sizes, which tend to increase in size and density as they approach the coating surface. In accordance with traditional bubble growth and nucleation theory, the bubble size generally increased with increasing temperature and was determined by the thermal budgets (product of temperature and irradiation duration), whereas bubble shapes generally changed from spherical to faceted when the bubbles grew [55, 56].

Obviously, in this work, the growth of the He bubbles in Cr layers primarily depends on the He concentration, as the apparent variation in bubble morphology arises from the depth-dependent He concentration gradient and proximity to CrAlSiN interfaces. Conversely, the bubbles in the CrAlSiN layers are too small to be discernible (**Fig. 4(e)**). Notably, these bubbles evolve into white dots of several nanometers in size along the interfaces (I) and ultimately disappear within the CrAlSiN layers. Additionally, in the CrAlSiN layers (**Fig. 4(f)-(g)**), some seemingly amorphous complexion films (light contrast) surround the nanograins (dark contrast). These amorphous intergranular films (AIFs), characterized by disordered atomic packing structures and low density, can trap and transport certain He atoms [57, 58]. Thus, it can be inferred that the He atoms/bubbles may migrate towards the opposite interfaces (II) when the AIFs connect to a network, *i.e.*, formation of nanochannels.

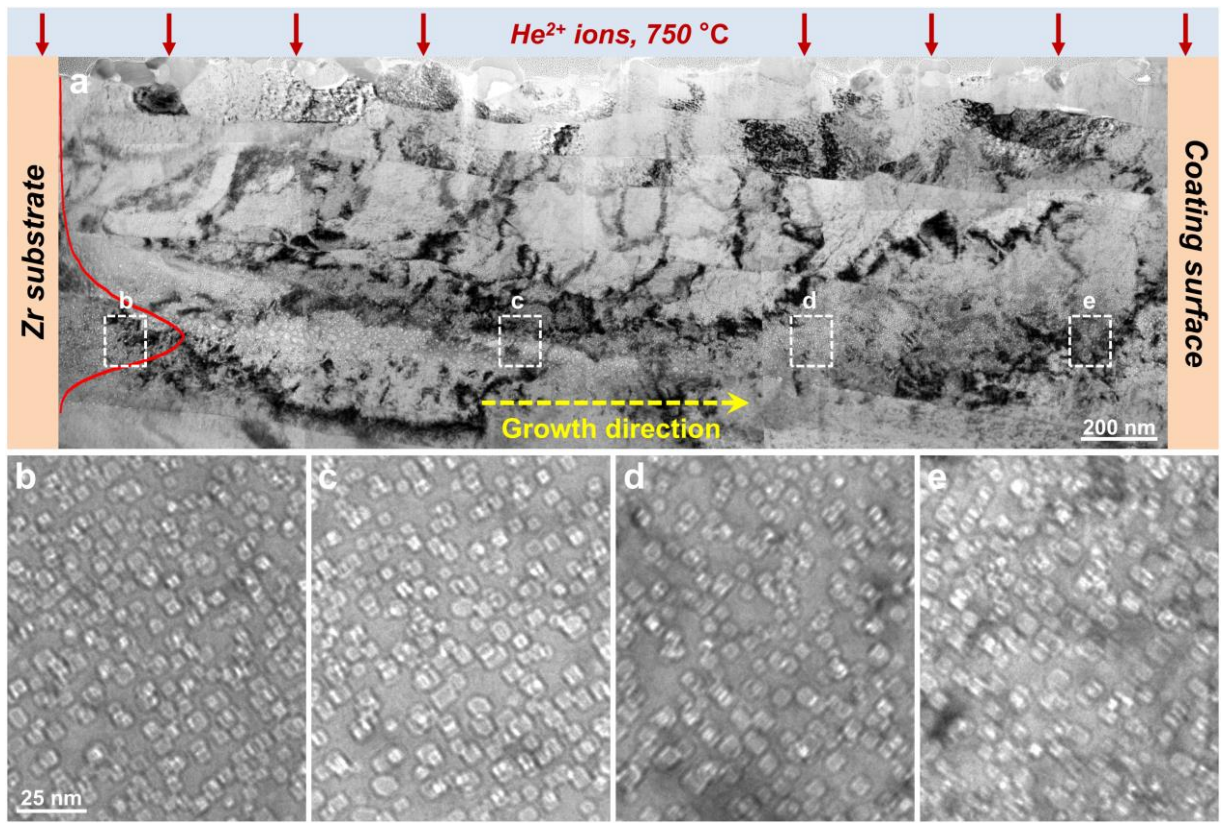

**Fig. 3** (a) Cross-sectional TEM images showing the characteristics of He bubbles in the 750 °C-irradiated Cr coating. (b)-(e) Enlarged TEM images of the areas marked in (a).

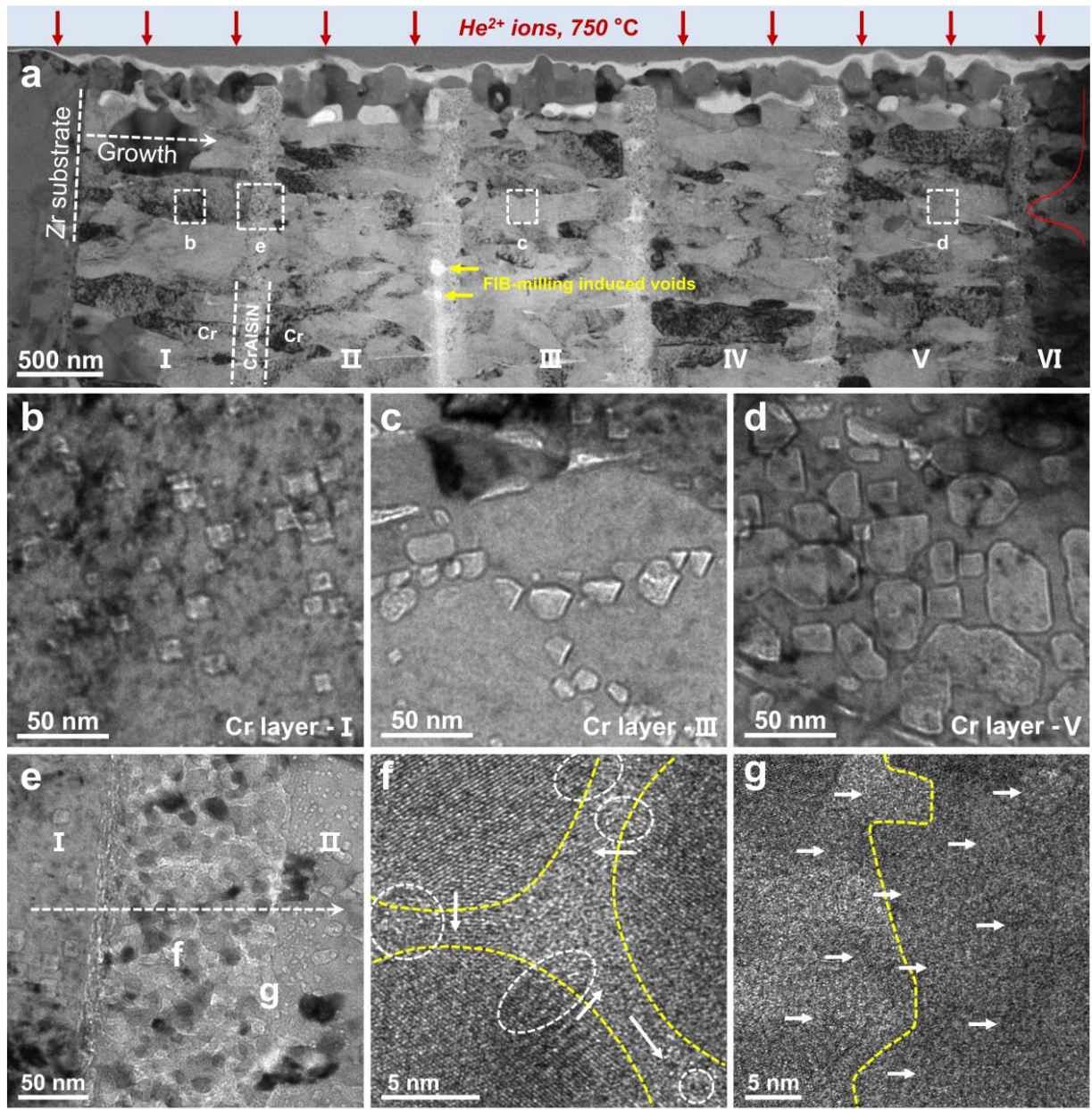

**Fig. 4** TEM characterization of the Cr/CrAlSiN multilayer coating (sample S1) after the $He^{2+}$ irradiation at 750 °C. (a) Low-magnification overview TEM image (different Cr layers are labeled with I, II, III, IV, V, and VI for distinction and reference). (b-e) Enlarged TEM images showing the distribution and morphology of He bubbles within the irradiated multilayer structure (more information can be found in Fig. S1 of the Supplementary Material). The Cr layers labeled I, III, and V correspond to those marked in (a). (f-g) High-resolution TEM (HRTEM) images of the selected areas in (e), where He bubbles are outlined by white dashed circles, and their potential migration directions are indicated by white arrows.

It should be emphasized that Fig. 4 provides a qualitative visualization of the spatial variation in He bubble morphology across the multilayer, whereas the trend of decreasing bubble diameter accompanied by increasing number density near CrAlSiN interfaces is quantitatively supported by the statistical analysis presented in Fig. 5. Therefore, the interpretation of bubble size–density gradients is based primarily on the quantitative data in Fig. 5, with Fig. 4 serving as a representative microstructural illustration. **Fig.5** quantifies the He bubble evolution in terms of diameter, density, and swelling in the Cr/CrAlSiN coating along the growth direction. Bubble diameter, surface coverage, density, and swelling were calculated using bright-field TEM images. Swelling ($S$) was determined via Eq. (1):

$$S = \frac{V_{bub}}{V_0 - V_{bub}} \times 100\% = \frac{\sum_{i=1}^{N} \frac{4\pi r_i^3}{3}}{A \times \delta - \sum_{i=1}^{N} \frac{4\pi r_i^3}{3}} \times 100\% \tag{1}$$

where $V_{bub}$ is the total bubble volume, and $V_0$ is the unirradiated material volume, $A$ is the TEM image area of layer $i$ and $\delta$ is the sample thickness. A periodic variation shows bubble diameter decreasing sharply near each CrAlSiN interface while density increases. For example, in the Cr layer-I, as it approaches the CrAlSiN layer, the average He bubble size decreases from ~7.40 nm to ~2.43 nm, while the bubble density increases from ~$1.1 \times 10^{22}$ m$^{-3}$ to ~$5.2 \times 10^{22}$ m$^{-3}$. Moreover, the He bubbles in the Cr layers exhibit a gradient distribution along the growth direction. The average bubble diameters in the six Cr layers are approximately 6.85 nm, 8.15 nm, 8.05 nm, 10.40 nm, 11.81 nm, and 11.88 nm, respectively, indicating a trend of increasing bubble size toward the coating surface. Notably, swelling also shows a gradual increase along the growth direction. The average swelling values in the six Cr layers are approximately 0.47%, 0.94%, 0.86%, 2.02%, 3.20%, and 6.50%, respectively.

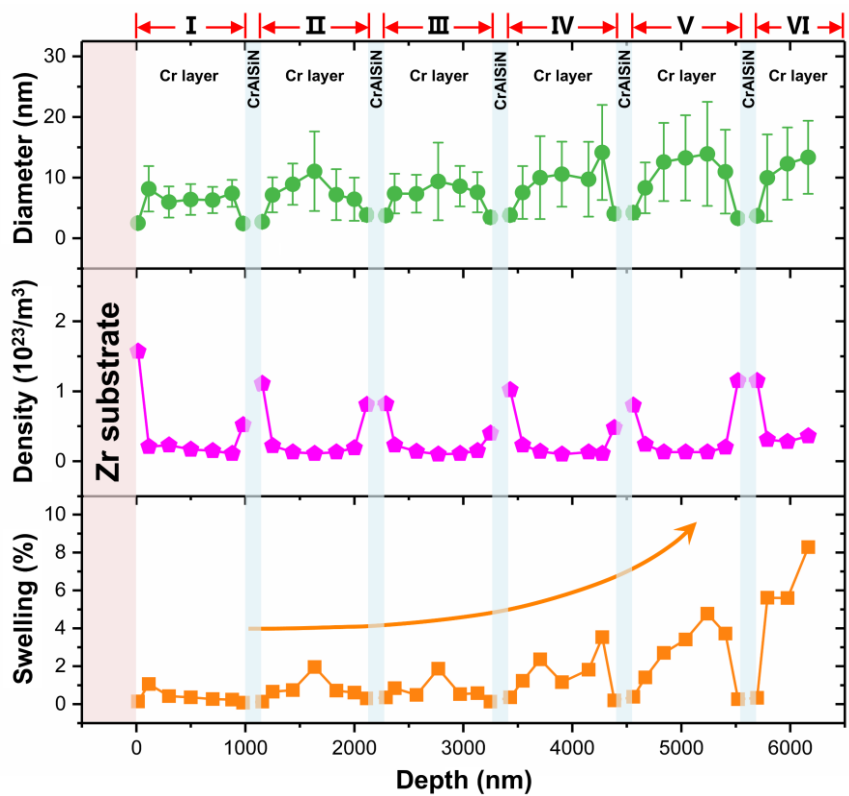

**Fig. 5** Depth dependences of bubble diameter, volume density, and cavity swelling at different depths in the Cr layers of the He-irradiated Cr/CrAlSiN coating (S1). The regions assigned with I, II, III, IV, V, and VI correspond to the Cr layers marked in **Fig. 4**. Each Cr layer was divided into seven regions for analysis.

*3.3    In-situ TEM observation of the He ions-irradiated coating during the annealing*

To further validate the proposed dynamic transport mechanism, *in-situ* TEM was employed to monitor the evolution of He bubbles in sample S2, which had been pre-irradiated at 500 °C. This observation was carried out during a controlled annealing process up to 750 °C, following the *in-situ* heating profile shown in **Fig. 2**. *In-situ* TEM micrographs of a selected Cr/CrAlSiN interface (locating the middle of **Fig. 2b**) are displayed in **Fig. 6(a)-(f)**, with four specific regions (Regions 1-4) chosen for detailed analysis of He bubble behavior. The corresponding quantitative data, bubble size, density, and swelling, are summarized in **Fig. 6(g)-(i)**. Regions 1-3, located within the Cr layer, contain numerous He bubbles with diameter sizes of ~2.5-3.0 nm. These bubble sizes remain relatively stable throughout the annealing process, with no significant coarsening or bubble aggregation observed up to 650 °C. However, both bubble density and associated swelling in these regions decrease with increasing temperature, indicating a possible depletion of He content. This effect is most prominent in Region 1, which is positioned closest to the Cr/CrAlSiN interface. At 750 °C, the bubble density and swelling in this region drop to approximately ~1.32 × 10$^{23}$ /m$^3$ and ~0.13%, representing a ~55% reduction compared to RT values (~2.92 × 10$^{23}$ /m$^3$, ~0.29%). The trend suggests outward migration of He atoms away from the Cr layer under thermal activation. In contrast, an opposite behavior is observed in Region 4, located within the CrAlSiN layer. Here, both the bubble density and swelling increase with annealing temperature: bubble density rises from 0.66 × 10$^{23}$ /m$^3$ at RT to 1.30 × 10$^{23}$ /m$^3$ at 750 °C, while swelling increases from 0.03% to 0.08%, correspondingly.

This indicates that the CrAlSiN layer acts as an effective sink or trap for migrating He atoms. Given its dense nanocomposite structure and low-defect network, the CrAlSiN layer appears to hinder further diffusion, thus accumulating He bubbles that have migrated from the neighboring Cr regions (Regions 1-3). These observations strongly support the hypothesis that the CrAlSiN layers function as dynamic "trapping stations" within the multilayer architecture, playing a critical role in mediating the redistribution and stabilization of He atoms during thermal treatment. Additionally, under cross-sectional in situ TEM conditions, He atoms redistribution during annealing toward the free surfaces of the thin lamella cannot be excluded, as demonstrated elsewhere [59]. However, in the present study, the observed bubble evolution is spatially heterogeneous and strongly correlated with the Cr/CrAlSiN interfaces, rather than being uniform across the foil thickness as would be expected for purely surface-driven helium loss. Moreover, bubble reduction in the vicinity of the interfaces is accompanied by bubble accumulation within adjacent layers, indicating helium redistribution and trapping at interfaces rather than simple escape through free surfaces. Taken together, these observations indicate that interface-mediated helium transport plays a dominant role in the observed behavior, while surface-related helium escape may occur as a secondary effect under thin-foil conditions.

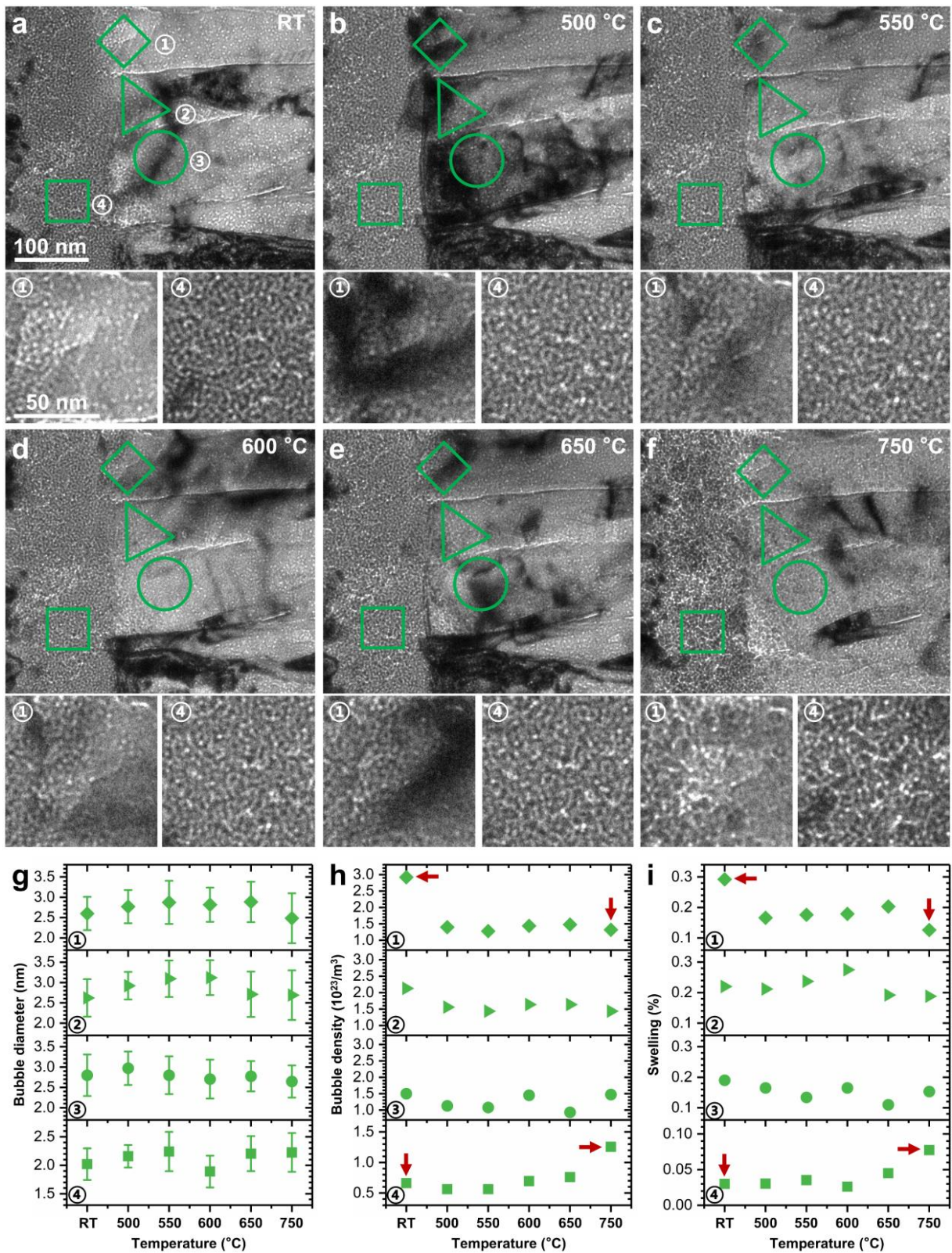

**Fig. 6.** (a-f) *In-situ* TEM images showing the evolution of He bubbles in the Cr/CrAlSiN coating (sample S2) under different temperatures. The images were taken around the peak damage region, encompassing a CrAlSiN layer (on the left), a Cr layer (on the right), and the interface between the two layers. (a) Room temperature (RT), (b) 500 °C, (c) 550 °C, (d) 600 °C, (e) 650 °C, (f) 750 °C. (g-i) Temperature dependences of (g) bubble diameter, (h) volume

density, and (i) cavity swelling at different regions marked in (a-f). The red arrows in (h) and (i) indicate the direction of apparent helium redistribution from the Cr layer to the CrAlSiN layer during annealing, highlighting the role of interfaces as preferential sinks.

*3.4. Microstructure of the coating/substrate interfaces after the He ions irradiation at 750 °C*

**Fig. 7** compares the irradiation response of the coating/substrate (C/S) interfaces in two systems, Cr-coated Zr alloy and Cr/CrAlSiN multilayer-coated Zr alloy, after 300 keV He ion irradiation at 750 °C. This setup allows direct observation of radiation-induced structural changes at the interface under realistic service-like conditions. As shown in Fig. 7(a), large defects with lateral dimensions approaching ~100 nm are observed on both sides of the Cr/Zr interface. These features are identified as vacancy-dominated voids formed at the interface, rather than as individual He bubbles. Their formation is attributed to a Kirkendall-type imbalance in atomic fluxes arising from the strong chemical driving force and asymmetric atomic mobilities of Cr and Zr under irradiation-assisted interdiffusion at elevated temperature. Smaller He bubbles are frequently observed attached to or decorating the surfaces of these interfacial voids. In this configuration, He acts primarily to stabilize existing vacancy clusters and void surfaces rather than to nucleate the large defects themselves. Although no extensive elemental interdiffusion is observed across the entire interface, the EDS inset reveals a partial overlap of Cr and Zr signals (several nanometers), specifically at the interfacial region. This indicates the presence of a thin reaction layer, most likely corresponding to the $ZrCr_2$ intermetallic phase [10], which is thermodynamically favored at elevated temperatures. The formation of such an interfacial phase may alter local defect sink behavior, facilitating the absorption of self-interstitial atoms at the interface and producing a surplus of vacancies. These excess vacancies then promote the accumulation and coalescence of He atoms into large bubbles and cavities. The resulting interfacial damage compromises both the mechanical integrity and thermal stability of the coating/substrate system.

By comparison, the Cr/CrAlSiN multilayer-coated sample shows a markedly different behavior. As presented in **Fig. 7(b)**, the C/S interface remains structurally intact and free from large cavities, even under identical irradiation conditions. N-enrichment is observed at the interface region, indicating radiation-induced migration of N atoms from the adjacent CrAlSiN layers toward the Zr substrate. These diffusing N atoms likely occupy radiation-induced vacancies and suppress vacancy clustering, thereby preventing He bubble nucleation and coalescence. It is worth noting that the N concentration exhibits a local decrease followed by re-enrichment in the vicinity of He bubbles. This behavior is attributed to radiation-enhanced diffusion coupled with defect–solute interactions, whereby vacancy-rich regions associated with He bubbles temporarily deplete nearby N, while simultaneously promoting N segregation toward defect-rich zones and interfaces. Such non-monotonic solute redistribution is characteristic of irradiation-driven defect fluxes. It is also noted that the nitrogen concentration exhibits a local decrease followed by re-enrichment in the vicinity of helium bubbles. This non-monotonic distribution is attributed to radiation-enhanced diffusion coupled with defect–solute interactions. During irradiation, vacancy-rich regions associated with He bubbles can temporarily deplete nearby N, while simultaneously promoting N segregation toward defect-rich zones and interfaces. Such behavior is characteristic of irradiation-driven defect fluxes and solute redistribution, rather than simple compositional inhomogeneity. The above vacancy passivation mechanism stabilizes the interface and preserves bonding strength, demonstrating the superior irradiation resistance of the Cr/CrAlSiN multilayer system compared to the single Cr coating. As no CrAlSiN layer was intentionally deposited between the Zr substrate and the Cr layer, the presence of nitrogen at the coating/substrate interface is attributed to radiation-enhanced diffusion of N atoms originating from the CrAlSiN layers. Under high-temperature irradiation, defect-assisted diffusion pathways, including vacancy clusters and dislocation networks, can significantly enhance long-range solute transport even across micrometer-scale Cr layers. This interpretation is proposed as a plausible mechanism and requires further experimental validation, particularly for specimens irradiated at lower temperatures or subjected to post-irradiation annealing.

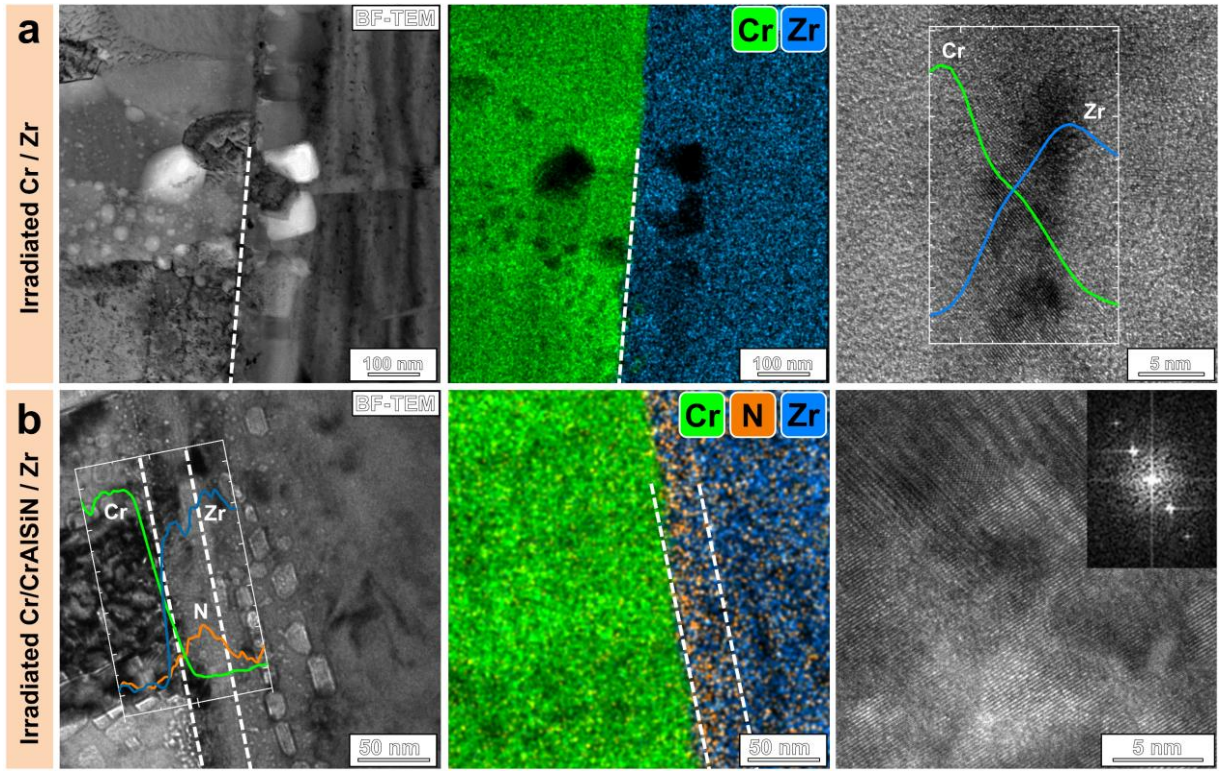

**Fig. 7** Cross-sectional TEM images and corresponding EDS elemental maps of the coating/substrate interfaces in (a) a Cr-coated Zr alloy and (b) a Cr/CrAlSiN multilayer-coated Zr alloy after 300 keV He ion irradiation at 750 °C. The C/S interface of Cr-coated sample in (a) showing the formation of large interfacial cavities and smaller helium bubbles on both sides of the interface. The large cavities are primarily associated with irradiation-assisted interdiffusion and asymmetric atomic fluxes across the Cr/Zr interface, leading to Kirkendall-type vacancy accumulation at elevated temperature. In contrast, the Cr/CrAlSiN-coated sample in (b) maintains an intact C/S interface without cavity formation, accompanied by N enrichment at the interfacial region.

## 4. Discussion

### *4.1 The dynamic transportation mechanism of Cr/CrAlSiN multilayered structure*

While the He fluence applied in this work exceeds levels expected in present-day light-water reactors, it provides insight into coating behavior under elevated gas concentrations that may become increasingly relevant for advanced reactor concepts operating at higher neutron fluxes or modified energy spectra. In this context, the present results are intended to elucidate fundamental mechanisms and upper-bound responses of candidate coating materials. The contrasting bubble morphologies observed in the Cr monolayer and the Cr/CrAlSiN multilayer under identical irradiation conditions (750 °C) highlight the decisive role of microstructural architecture in defect evolution. In the monolayer Cr coating, He bubbles exhibit uniform size and distribution across the damage region, with no evidence of bubble-denuded zones (**Fig. 3**). This uniformity indicates that the concentration of implanted He exceeds what can be accommodated by conventional defect sinks such as vacancy clusters [60], stacking faults [61], and grain boundaries (GBs) [62]. Although high-angle GBs can trap more He atoms due to their excess free volume [35], their scarcity in coarse-grained Cr limits overall storage capacity. Furthermore, the relatively large grain/column size increases the diffusion distance needed for He atoms and point defects to reach annihilation sites, reducing sink efficiency compared to fine-grained or nanostructured materials [63, 64]. In contrast, periodic insertion of CrAlSiN layers fundamentally alters the spatial evolution of bubbles in the multilayered structure. As shown in **Figs. 4-5**, He bubbles display a periodic modulation in both diameter and density along the growth direction. Near each CrAlSiN interface, bubble diameter decreases sharply while number density increases, suggesting active He fragmentation and reorganization.

The CrAlSiN layers contain uniformly distributed, low-density nanochannels, which serve as efficient sinks and transformation sites. Previous work on similar CrAlSiN coatings confirmed that these nanochannels prevent bubble coalescence and irradiation hardening, unlike Cr and CrAlSi coatings that show pronounced bubble aggregation and Kirkendall porosity under equivalent conditions [16]. These channels and amorphous-like regions promote the transformation of spherical bubbles into platelets, which are thermodynamically more efficient helium storage configurations, capable of storing approximately three times more He atoms per unit volume than spherical bubbles [65]. Their fragmentation reduces local swelling and delays growth and linkage into larger cavities. The periodic CrAlSiN layers therefore act as "mesh filters" (schematically illustrated in **Fig. 8**), intercepting migrating He atoms/bubbles from adjacent Cr layers and breaking them into smaller platelets. In the schematic illustration shown in Fig. 8, the 'left' and 'right' sides are defined solely for clarity to represent regions of relatively larger and smaller He bubbles, respectively, rather than absolute spatial directions. The schematic is intended to illustrate the conceptual migration of helium from regions with larger bubbles toward Cr/CrAlSiN interfaces, where He is redistributed into a higher density of smaller bubbles or atomic-scale clusters.

In-situ TEM annealing further confirms directional migration of He from Cr into CrAlSiN layers, accompanied by a reduction in bubble density in Cr and increased trapping within CrAlSiN (**Fig. 6**). These observations collectively demonstrate that He redistribution in the multilayer is governed by interface-assisted transport rather than accumulation within the Cr matrix. This hierarchical He accommodation mechanism, driven by elastic mismatch, chemical heterogeneity, and engineered free volume, significantly suppresses bubble coalescence and swelling. Unlike conventional void-engineered or multilayer systems (e.g., Cu/Nb [66]), the CrAlSiN layers enable repeated fragmentation and release of He toward the next Cr layer, rather than trapping it permanently. The result is a spatially modulated, non-coarsening bubble distribution that improves irradiation tolerance beyond what monolithic Cr can achieve. These results confirm that the multilayer's function is not merely passive defect accommodation but active modulation of He transport through periodic, nanoengineered sinks that redistribute and reshape bubbles before coalescence can occur.

Furthermore, atomistic simulations provide insight into the role of nitrogen in defect stabilization within CrAlSiN [16]. Density functional theory calculations have shown that the removal of Al or Si atoms from the CrAlSiN structure, creating vacancy-like free volumes, induces a pronounced rearrangement of surrounding N atoms. This rearrangement leads to the formation of new N–metal bonds, effectively reconstructing the local bonding environment and annihilating the free volume rather than allowing vacancy stabilization. Such behavior suppresses vacancy clustering and limits the formation of stable nucleation sites for helium bubbles. Consequently, nitrogen imparts a "self-healing" character to CrAlSiN by actively counteracting radiation-induced free volume defects. In addition, the formation of chemically and structurally distinct interfacial phases modifies local defect sink behavior. Interfaces are known to preferentially absorb self-interstitial atoms due to their reduced formation energies relative to the bulk, leading to enhanced interstitial annihilation and a local vacancy surplus in adjacent regions [32, 35, 49]. When combined with nitrogen-mediated vacancy stabilization, this interfacial sink effect promotes defect recombination and suppresses the growth and coalescence of helium bubbles. This synergistic interaction between interface sink strength and nitrogen chemistry provides a mechanistic basis for the improved radiation tolerance observed in the Cr/CrAlSiN multilayer coating.

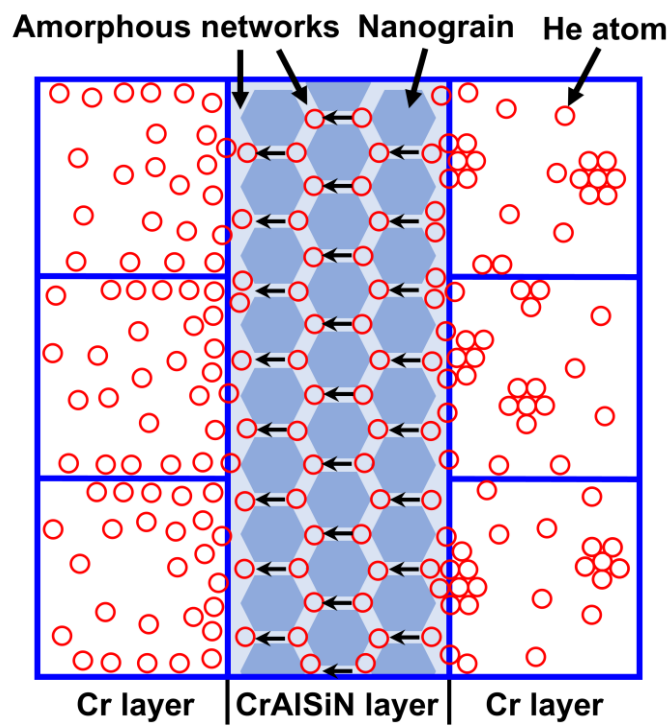

**Fig. 8** The schematic drawing of the absorption and diffusion mechanism of He atoms in the Cr/CrAlSiN coating (supposing that the He migration direction is from right to left).

### *4.2  The irradiation stability of coating-substrate interfaces*

The irradiation stability of the Cr/Zr interface is critically important for the performance of Cr-coated zirconium alloy claddings in nuclear applications. Previous studies have shown that irradiation with 20 MeV $Kr^{8+}$ ions at 400 °C (to a damage level of ~10 dpa) leads to the formation of a $Zr(Cr,Fe)_2$ intermetallic phase at the Cr/Zr interface, featuring good atomic matching and crystallographic compatibility [10]. However, when the irradiation dose was increased to 20-35 dpa using 5 MeV Au$^{5+}$ ions at 400 °C, this interphase underwent decomposition, accompanied by a "sharpening" effect linked to continuous intermixing and phase instability [67]. In the present work, both large interfacial cavities and He bubbles were observed on either side of the Cr/Zr interface (**Fig. 7a**). This behavior is primarily associated with the elevated irradiation temperature (750 °C), at which thermally activated, irradiation-assisted interdiffusion across the Cr/Zr interface becomes significant. Due to the strong chemical driving force and asymmetric atomic mobilities of Cr and Zr, a Kirkendall-type imbalance in atomic fluxes develops, leading to the accumulation of vacancies at the interface. These vacancy-rich regions provide favorable sites for cavity nucleation, independent of the vacancy production rate associated with He implantation. Once formed, such cavities can effectively trap mobile He atoms and He–vacancy complexes, promoting He stabilization and cavity growth. Consequently, He acts mainly as a cavity-stabilizing species rather than the primary source of vacancies. Large interfacial cavities, often exceeding 100 nm in size, therefore result from the combined effects of irradiation-assisted interdiffusion, phase instability at the Cr/Zr interface, and He trapping. This mechanism is consistent with established models of Kirkendall-type cavity formation under high-temperature irradiation conditions. These results indicate that Cr/Zr interface stability degrades significantly at elevated irradiation temperatures (e.g. above ~750 °C), even though earlier work has reported stable Cr/Zr interfaces under room-temperature or low-temperature irradiation. At these high temperatures, mutual interdiffusion between Cr and Zr leads to the formation of an intermixed Cr-Zr layer, commonly composed of ZrCr$_2$ Laves phase. This phase is prone to cracking, delamination, and mechanical incompatibility under stress. Additionally, Cr can dissolve into β-Zr at high temperatures [68], accelerating coating consumption and promoting chemically driven anisotropic phase evolution in Zircaloy.

In contrast, the multilayered Cr/CrAlSiN coating exhibits a substantially more stable C/S

interface under identical irradiation conditions. No large cavities or interfacial bubbles were detected (**Fig. 7(b)**). Instead, N-enrichment was detected at the interface, indicating outward migration of N from the CrAlSiN layers into the Zr substrate region. The formation of this N-enriched Zr(N) interlayer is consistent with the strong thermodynamic driving force for Zr-N bonding ($\Delta G^f$ (ZrN) ≈ −262.8 kJ/mol at ~800 °C). Similar N diffusion into Zr has been reported in Cr-CrN-Cr multilayered coatings, where the resulting Zr(N) layer effectively blocks mutual atomic interdiffusion [46]. The presence of N at the interface has multiple irradiation-stabilizing effects. Firstly, N and Zr can form strong covalent bonds, reducing the escape of Zr atoms from lattices and forming vacancies. Moreover, as an interstitial solute with a small atomic radius (~56 pm), N can more readily occupy vacancies and exhibit a strong affinity for vacancy-like free volumes, where they undergo local rearrangement and form new directional bonds with surrounding metal atoms [16]. This bonding reconstruction effectively annihilates or stabilizes free volume and the energetic barrier for vacancy clustering and void nucleation, therefore suppressing vacancy clustering cavity formation at the interface. Additionally, the incorporation of N perturbs local bonding and reduces the mobility of both long-range diffusing species and He atoms. This behavior parallels observations in high-entropy alloys and N-stabilized systems, where chemical heterogeneity and solute-induced lattice strain inhibit defect transport and facilitate point defect recombination [69]. Notably, in the Cr/CrAlSiN multilayer, the Zr(N) interlayer contains no visible He bubbles, unlike the adjacent Cr and Zr regions, indicating that He is either excluded from or immobilized within the interface. In addition to Cr and N, alloying elements such as Al and Si in the CrAlSiN layers may also influence helium behavior indirectly. Al and Si are known to contribute to the stabilization of amorphous or nanocomposite CrAlSiN structures by increasing chemical disorder and disrupting long-range crystallographic order. This structural stabilization can reduce long-range atomic mobility and modify defect formation and migration energies, thereby indirectly affecting helium transport and bubble evolution. However, within the spatial resolution of EDS, no distinct Al- or Si-rich segregation zones were detected around He bubbles, indicating that their influence is primarily indirect rather than through direct helium trapping. The Zr substrate mainly serves as a mechanical and chemical support, and no significant Zr diffusion into the coating or preferential interaction with helium was observed under the present irradiation conditions.

Comparing **Fig. 7a** and **Fig. 7b**, it is evident that the Cr monolayer coating undergoes severe interfacial degradation via vacancy-driven cavity formation, cavity growth is governed by irradiation-assisted interdiffusion and vacancy stabilization at the interface, with He acting to stabilize and enlarge existing vacancy clusters. It is emphasized that the observed cavity formation does not require a high bulk vacancy production rate, but instead originates from interface-specific diffusion asymmetry and phase instability under high-temperature irradiation. While the Cr/CrAlSiN multilayer resists both chemical and mechanical destabilization. The N-enriched interlayer acts as a reactive diffusion barrier and a defect passivation zone, maintaining interfacial cohesion even at elevated temperatures. These findings underscore the importance of coupling interface chemistry with architecture-driven He management. While electroplated Cr coatings are industrially attractive, their limited defect sink capacity and susceptibility to interdiffusion compromise high-temperature irradiation performance. In contrast, the Cr/CrAlSiN multilayer system introduces spatially distributed defect sinks, suppresses large bubble formation, and enables dynamic redistribution of He while simultaneously stabilizing the C/S interface through N incorporation. Overall, the contrast between the two systems demonstrates that chemical stabilization of buried interfaces, coupled with nanoscale architectural control, is essential for suppressing cavity formation and ensuring interfacial integrity under high-temperature irradiation.

5. **Conclusion**

Cross-sectional He irradiation provides a unique perspective on the irradiation response of Cr-based coatings by allowing direct observation of buried interface behavior and defect evolution across multiple layers. In this study, the approach revealed distinct differences in helium bubble dynamics between Cr monolayers and Cr/CrAlSiN multilayers deposited on Zr substrates. At 750 °C, the Cr monolayer exhibited homogeneous bubble formation, significant interdiffusion, and large Kirkendall-type cavities at the Cr/Zr interface, indicating poor interfacial stability. In contrast, the Cr/CrAlSiN multilayer displayed spatial modulation of bubble morphology and density, with nanoscale helium platelets forming at CrAlSiN interfaces. A nitrogen-enriched Zr(N) layer formed in situ at the CrAlSiN/Zr interface, serving as an effective diffusion barrier and suppressing both cavity formation and interfacial degradation. The nanochannel architecture of the multilayer further enhanced dynamic defect accommodation and helium trapping, reducing bubble coalescence and maintaining coating integrity. These results confirm that interface and chemistry engineering, through nitrogen incorporation and periodic multilayer structuring, can dramatically improve irradiation resistance and interfacial stability in Cr-based coatings. The cross-sectional irradiation methodology used here provides a robust experimental framework for understanding helium–defect–interface interactions in complex multilayers and may guide the design of radiation-tolerant coatings for next-generation nuclear energy systems. The results highlight the critical synergy between interfaces and chemical complexity in governing He behavior, providing mechanistic guidance for the design of advanced accident-tolerant cladding coatings rather than direct lifetime prediction. Certainly, dual-beam irradiation studies will be needed in the future to replicate the actual service conditions more accurately.

**Acknowledgments**

This research is supported by the National Natural Science Foundation of China (No. 12575304, 12575305) and the International Sci-tech Cooperation Projects of Ningbo City (2024H026, 2024J072), and. The authors are grateful for ion irradiation experiments to the staff of the 320 kV Multidisciplinary Research Platform and the accelerator crew of the Heavy Ion Research Facility in Lanzhou (HIRFL). This work was financially supported by the European Union under the project Robotics and advanced industrial production (Reg. No.

CZ.02.01.01/00/22_008/0004590). CzechNanoLab project LM2023051, funded by MEYS CR, is gratefully acknowledged for the financial support of the measurements/sample fabrication at LNSM Research Infrastructure.